\documentclass[12pt,superscriptaddress]{article}
\pdfoutput=1
\usepackage{jheppub}
\usepackage{graphicx}  
\usepackage{dcolumn}   
\usepackage{bm}        
\usepackage{amssymb}   
\usepackage{slashed} 

\newcommand{\beq}{\begin{equation}}
\newcommand{\eeq}[1]{\label{#1}\end{equation}}
\def\beqa{\begin{eqnarray}}
\def\eeqa#1{\label{#1}\end{eqnarray}}
\newcommand{\eeqn}{\end{equation}}
\newcommand{\CR}{\notag \\}
\newcommand{\leqn}[1]{(\ref{#1})}

\def\mg{m_{\tilde{g}}}
\def\mt{m_{\tilde{t}}}
\def\go{\tilde{g}}
\def\st{\tilde{t}}

\def\met{E_T^{\rm miss}}
\def\GeV{~{\rm GeV}}

\def\stacksymbols #1#2#3#4{\def\theguybelow{#2}
    \def\vp{\lower#3pt}
    \def\sp{\baselineskip0pt\lineskip#4pt}
    \mathrel{\mathpalette\intermediary#1}}

\def\intermediary#1#2{\vp\vbox{\sp
     \everycr={}\tabskip0pt
     \halign{$\mathsurround0pt#1\hfil##\hfil$\crcr#2\crcr
              \theguybelow\crcr}}}

\def\lsim{\stacksymbols{<}{\sim}{2.5}{.2}}
\begin{document}


\title{
The Same-Sign Dilepton Signature of RPV/MFV SUSY}

\author[a]{Joshua Berger} 
\author[b]{, Maxim Perelstein}
\author[b]{, Michael Saelim}
\author[b]{and Philip Tanedo}

\affiliation[a]{SLAC National Accelerator Laboratory, 2575 Sand Hill, Menlo Park, CA 94025}

\affiliation[b]{Laboratory for Elementary Particle Physics, 
	     Cornell University, Ithaca, NY 14853, USA}
	     
\emailAdd{jberger@slac.stanford.edu}     
\emailAdd{mp325@cornell.edu}     
\emailAdd{mjs496@cornell.edu}     
\emailAdd{pt267@cornell.edu}     

\abstract{The lack of observation of superpartners at the Large Hadron Collider so far has led to a renewed interest in supersymmetric models with R-parity violation (RPV). In particular, imposing the Minimal Flavor Violation (MFV) hypothesis on a general RPV model leads to a realistic and predictive framework. Naturalness suggests that stops and gluinos should appear at or below the TeV mass scale. We consider a simplified model with these two particles and MFV couplings. The model predicts a significant rate of events with same-sign dileptons and $b$-jets. We re-analyze a recent CMS search in this channel and show that the current lower bound on the gluino mass is about 800 GeV at 95\% confidence level, with only a weak dependence on the stop mass as long as the gluino can decay to an on-shell top-stop pair. We also discuss how this search can be further optimized for the RPV/MFV scenario, using the fact that MFV stop decays often result in jets with large invariant mass. With the proposed improvements, we estimate that gluino masses of up to about $1.4$ TeV can be probed at the 14 TeV LHC with a 100 fb$^{-1}$ data set.}


\maketitle

\newpage

\section{Introduction}

Supersymmetry (SUSY) remains one of the most compelling ideas for extending the Standard Model (SM). While SUSY is clearly broken in nature, naturalness of electroweak symmetry breaking strongly suggests that it should be restored at an energy scale $\lsim 1$ TeV. This would require the SUSY partners of the SM particles to appear at that scale. However, experiments conducted in 2010--2012 at the Large Hadron Collider (LHC) have seen no evidence for such superpartners, placing lower bounds on the masses of some of them, squarks and gluinos, well in excess of 1 TeV. This apparent contradiction led many theorists to question the assumptions underlying the LHC searches. One of the most important assumptions is R-parity conservation, which implies that the lightest superpartner (LSP) is stable. A stable LSP in turn implies that each event with superpartner production contains either missing transverse energy (MET) or exotic charged tracks, either of which provides a good handle to distinguish such events from the SM backgrounds. Most LHC searches make extensive use of such handles. If there is no conserved R-parity, these searches would not be applicable and the LHC bounds would be weakened significantly, removing conflict with naturalness.

From the theoretical point of view, R-parity is {\it not} required by SUSY: it is an additional discrete symmetry. The motivation for introducing this extra symmetry is purely phenomenological: it forbids baryon (B) and lepton (L) number violating operators that would otherwise induce rapid proton decay. However, proton decay and other tightly constrained B- and L-violating processes may be forbidden or suppressed to acceptable levels {\it without} introducing R-parity. An interesting proposal along these lines has been made recently by Csaki, Grossman and Heidenreich~\cite{CGH} (see also~\cite{NS}). The authors start with a minimal SUSY model without R-parity. They then impose the Minimal Flavor Violation (MFV) hypothesis, which is strongly motivated by flavor physics constraints on SUSY, on the full superpotential, including B- and L-violating operators. The MFV hypothesis in effect imposes an accidental approximate R-parity on the first two generations and greatly suppresses dangerous operators such as those that induce proton decay. At the same time, there are non-trivial R-parity violating (RPV) couplings involving the third generation which are sufficient to render the LSP unstable on collider time scales and weaken the LHC bounds. This is the framework that we focus on in this paper.\footnote{For recent work on complete SUSY models realizing this framework, see Refs.~\cite{UVcompletions}.}

As for any SUSY model, the collider phenomenology of MFV SUSY depends sensitively on the superpartner spectrum. This, in turn, is determined by the details of the SUSY breaking sector and mediation, for which many possible models have been proposed. In this paper, we focus on a simple scenario motivated by bottom-up naturalness considerations. It is well known that the only superpartners required to be light ($\lsim 1$ TeV) by naturalness are the stops $\tilde{t}_{1,2}$, the Higgsino $\tilde{H}$, and the gluino $\tilde{g}$: see, for example, Ref.~\cite{Raman} for a clear and careful explanation of this point. Of these, $\tilde{H}$ 
has a suppressed production rate due to its weak coupling. Thus, it will not have a considerable impact on phenomenology as long as it is not the LSP. We will therefore consider a simplified model~\cite{Simpl} with just two states: a gluino $\tilde{g}$ and a stop $\tilde{t}$. All other SUSY particles are assumed to be either too heavy or too weakly coupled to be relevant at the LHC.\footnote{We do not include a left-handed sbottom $\tilde{b}_L$ in our simplified model even though its presence at the same mass scale as the stop is well motivated. In MFV SUSY, the dominant sbottom decays typically involve the top quark, $\tilde{b}\to t c$ or $\tilde{b}\to t \tilde{\chi}^-$, so that gluino cascades via sbottoms can still produce the same-sign dilepton signature. Thus we expect that the bounds derived here would qualitatively apply to most MFV SUSY models with $\mg>m_{\tilde{b}}$ as well.} We assume that the stop is the LSP, as motivated by naturalness considerations, and that $m_{\tilde{g}}>m_{\tilde{t}}+m_t$. We focus on gluino pair-production, $pp\to \tilde{g}\tilde{g}$, followed by a cascade decay:
\beqa
\go&\to& \st \bar{t},~~~\st \to \bar{b} \bar{s} \CR
& & {\rm or} \CR
\go&\to& \st^\ast t,~~~\st^\ast \to bs\,.
\eeqa{cascade} 
The branching ratio for each of these channels is 50\%, assuming a purely Majorana gluino. With probability of 50\%, the gluino pair will produce a same-sign top pair ($tt$ or $\bar{t}\bar{t}$). If each top decays leptonically, the final state will contain two same-sign leptons: $e^\pm e^\pm$, $\mu^\pm \mu^\pm$, or $e^\pm \mu^\pm$. Such ``same-sign dilepton" (SSDL) events are very rare in the SM, and the SSDL signature already plays a prominent role in the LHC SUSY searches. Typically, these searches demand substantial MET in addition to SSDL, reducing their sensitivity to the RPV cascades~\leqn{cascade} where the only sources of MET are neutrinos from leptonic top decays. However, the SSDL signature by itself is so striking that searches may be conducted even with no (or very low) MET cut, making them sensitive to RPV SUSY~\cite{AG,mesino,Asano,Harvard}.\footnote{Other signatures of RPV SUSY with light stops and gluinos have been discussed in Refs.~\cite{others1,others2}. SSDL signature from resonant slepton production has been discussed in~\cite{Dreiner}.} The first goal of this paper is to estimate the current bounds on our simplified model using the latest publicly available CMS search for the SSDL signature~\cite{CMS3}. This search uses $10.5$ fb$^{-1}$ of data collected at $\sqrt{s}=8$ TeV in the 2012 LHC run.

While the current SSDL searches already place interesting bounds on RPV SUSY, they are not optimized for this class of models. The second goal of this paper is to suggest ideas for optimizing this search that may be implemented by the experiments in the future. SSDL events in RPV SUSY have at least 6 parton-level jets. This high jet multiplicity can, by itself, provide an additional handle to suppress backgrounds. Moreover, two pairs of these jets come from stop decays. Depending on the gluino and stop masses, two regimes are possible. If $m_{\tilde{g}}-m_{\tilde{t}} \sim m_t$, the stops are typically non-relativistic in the lab frame and the two jets are well separated. In this regime, one simply needs to look for a resonance in the dijet invariant mass. The case 
$m_{\tilde{g}} \gg m_{\tilde{t}}$ is more interesting. In this case, the stops are predominantly relativistic, and their decay products are boosted in the direction of their motion. The two parton showers would typically be merged in a single jet, and the signatures of their ``stoppy" origin are hidden in the {\it substructure} of the jet. Recently, much work has been done on exploring observables sensitive to jet substructure (for a review, see~\cite{JSSreview}). We will show how some of these techniques can be used to further enhance the sensitivity of the SSDL search for RPV SUSY.

The rest of the paper is organized as follows. The current bounds on RPV SUSY derived from the recently published CMS search in the SSDL channel are presented in Section~\ref{sec:recast}. Additional cuts that can be used to improve the sensitivity of this search specifically in the RPV SUSY case are discussed in Section~\ref{sec:future}. Section~\ref{sec:conc} contains brief conclusions and outlook, while some of the details of the procedure used to recast the CMS search are presented in Appendix~\ref{app:recast}.

\section{Current Bounds: Recasting the CMS SSDL Search}
\label{sec:recast}

Both CMS and ATLAS perform searches for the SSDL signature, accompanied by MET and jets (with or without $b$-tag requirement), as part of their standard search strategy to look for R-parity conserving (RPC) SUSY with light gluinos and stops. These analyses have non-trivial sensitivity to the RPV SUSY cascade~\leqn{cascade} since leptonic top decays contain neutrinos which provide genuine MET, typically in the few tens of GeV range. While most RPC SUSY searches must impose a MET cut of at least 100 GeV to suppress SM backgrounds, the SSDL signature by itself is very rare in the SM so that such a strong MET cut is not required. The CMS collaboration recently published bounds based on a number of signal regions (SRs) with either no MET cut or sufficiently low  MET cuts (30--50 GeV) that are easily exceeded by the top-induced MET~\cite{CMS3}. While the CMS paper interprets the results in terms of RPC SUSY, it is straightforward to ``recast" their published data to provide limits on the RPV case.\footnote{Previous recasts of the LHC SSDL searches in terms of RPV SUSY have appeared in~\cite{AG,Harvard}. These searches use smaller data sets than the one considered here.}

	\begin{table}[t]
		\tabcolsep 2.7pt
		\begin{scriptsize}
			\begin{tabular}{|l|c|c|c|c|c|c|c|c|c|}
			\hline
                                               & SR0                    & SR1                    & SR2                    & SR3                    & SR4                    & SR5                    & SR6                    & SR7                    & SR8            \\
                        \hline
			No.\ of jets            & $\geq 2$               & $\geq 2$               & $\geq 2$               & $\geq 4$               & $\geq 4$               & $\geq 4$               & $\geq 4$               & $\geq 3$               & $\geq 4$       \\
			No.\ of $b$-tags           & $\geq 2$               & $\geq 2$               & $\geq 2$               & $\geq 2$               & $\geq 2$               & $\geq 2$               & $\geq 2$               & $\geq 3$               & $\geq 2$       \\
			$\ell$ charges         & $++/--$                & $++/--$                & $++$                   & $++/--$                & $++/--$                & $++/--$                & $++/--$                & $++/--$                & $++/--$        \\
			$\met$                 & ${>}0$\GeV             & ${>}30$\GeV            & ${>}30$\GeV            & ${>}120$\GeV           & ${>}50$\GeV            & ${>}50$\GeV            & ${>}120$\GeV           & ${>}50$\GeV            & ${>}0$\GeV     \\
			$H_T$                  & ${>}80$\GeV            & ${>}80$\GeV            & ${>}80$\GeV            & ${>}200$\GeV           & ${>}200$\GeV           & ${>}320$\GeV           & ${>}320$\GeV           & ${>}200$\GeV           & ${>}320$\GeV   \\
			\hline
			\end{tabular}
		\end{scriptsize}
\caption{Event characteristics required in the 9 signal regions (SRs) used in the CMS SSDL+MET+$b$ analysis~\cite{CMS3}. Note that the number of jets on the first line of
		the table includes both $b$-tagged and non-$b$-tagged jets. For the predicted background rates and the observed rates in each region, see Table 2 of Ref.~\cite{CMS3}.}
\label{tab:fromCMS}
	\end{table}

The cuts imposed by the CMS analysis are summarized in Table~\ref{tab:fromCMS}. The acceptance cuts are $p_T>40$ GeV, $|\eta|<2.4$ for jets (both $b$-tagged and non-$b$-tagged), and $p_T>20$ GeV, $|\eta|<2.4$
for electrons and muons. 
Events with a third lepton are vetoed if they contain an opposite-sign lepton pair with invariant mass below 12 GeV, or between 76 and 106 GeV, to avoid contamination from $Z$ decays. For more details on the CMS analysis, see Ref.~\cite{CMS3}. 

\begin{figure}[tb]
\begin{center}
\centerline {
\includegraphics[width=4in]{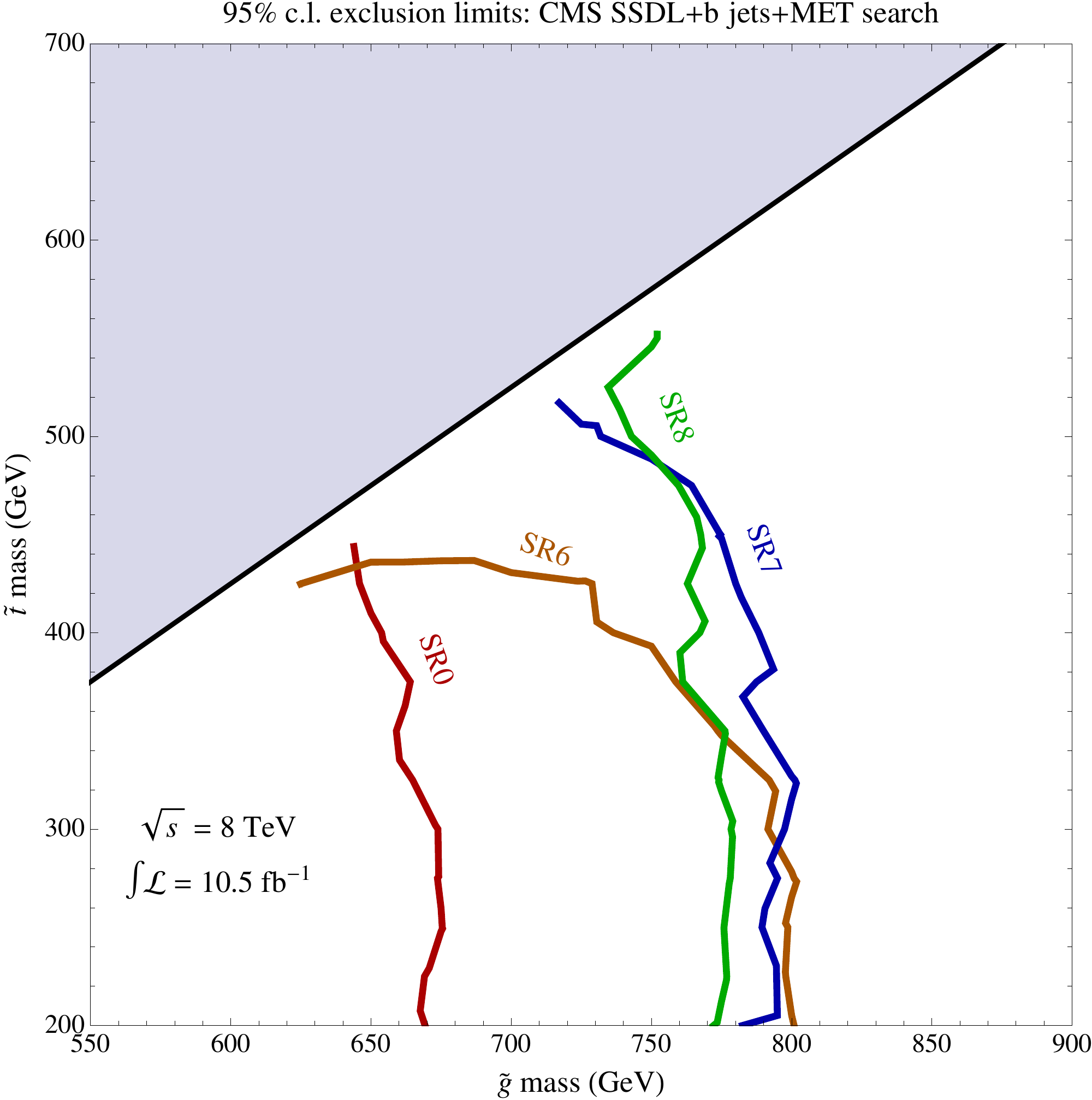}
}
\caption{95\%~CL exclusion of the RPV SUSY simplified model parameter space, based on the 4 most sensitive search regions (SRs) from the CMS SSDL+MET+$b$ search~\cite{CMS3} with $10.5$ fb$^{-1}$ of data collected at the 8 TeV LHC.}
\label{fig:bounds}
\end{center}
\end{figure}

In all nine signal regions, the data is consistent with the SM expectation, so an upper bound on the number of signal events can be set. We simulated the process $pp\to\tilde{g}\tilde{g}$, followed by the decays~\leqn{cascade} and the leptonic top decay on both sides, using {\tt Pythia 8.162}~\cite{pythia}, for a large set of $(\mg, \mt)$ points. The leading order (LO) cross section provided by {\tt Pythia} is multiplied by the NLO K-factor computed with {\tt Prospino 2.1}~\cite{prospino} for normalization. To compute the efficiency of the CMS cuts on the signal, we essentially follow the procedure described in the CMS report~\cite{CMS3} and its predecessors~\cite{SUS-11-020,SUS-12-017}. For details, see Appendix~\ref{app:recast}. The only non-trivial deviation from the CMS prescriptions concerns the treatment of lepton selection efficiencies. These have two factors: identification (ID) efficiency and the efficiency of the lepton isolation cut. CMS only published the combined lepton selection efficiency for a benchmark RPC SUSY point LM9~\cite{LM9}. However, the RPV SUSY signal is expected to have a significantly different lepton isolation efficiency: there is more hadronic activity, and, in some parts of the parameter space, the tops are boosted, resulting in a $b$-jet in close proximity to the lepton. To take this into account, we estimate the lepton isolation cut efficiency from our signal MC, at each $(\mg, \mt)$ point, and multiply by the lepton ID efficiency estimated by a separate simulation of the LM9 RPC SUSY signal. The cross section, acceptance and efficiency are then used to compute the number of expected signal events at each~$(\mg, \mt)$ point. Comparing this number with the background prediction and data provided by CMS and using the $CL_s$ method~\cite{CLs} yields the expected 95\% confidence level (CL) exclusion.

The results of this analysis are summarized by Fig.~\ref{fig:bounds}, which shows the 95\%~CL exclusion contours from the four most sensitive signal regions. We conclude that the current bound on the gluino mass is about 800 GeV. The bound is approximately independent of the stop mass as long as an on-shell decay $\tilde{g}\to\tilde{t} t$ is kinematically allowed. Note that this bound is somewhat stronger than the bound recently obtained in Ref.~\cite{Harvard} by recasting the ATLAS SSDL+MET+$j$ search~\cite{ATLAS}. The difference is especially pronounced in the region of relatively small gluino/stop mass splitting, where the ATLAS analysis loses sensitivity due to the large MET required ($\geq 150$ GeV). The remaining differences are accounted for by the slightly higher integrated luminosity of the CMS search, as well as the additional requirement of $b$-tagged jets imposed by CMS.

\section{Future Searches: Optimizing for the RPV}
\label{sec:future}

While the current SSDL+MET+$b$ searches already provide meaningful bounds on RPV SUSY, they are clearly not optimized for this model. In this section, we suggest ways to enhance their sensitivity to the RPV model, and demonstrate the improvements with a Monte Carlo analysis. 

\begin{figure}[t!]
\begin{center}
\centerline {
\includegraphics[width=4in]{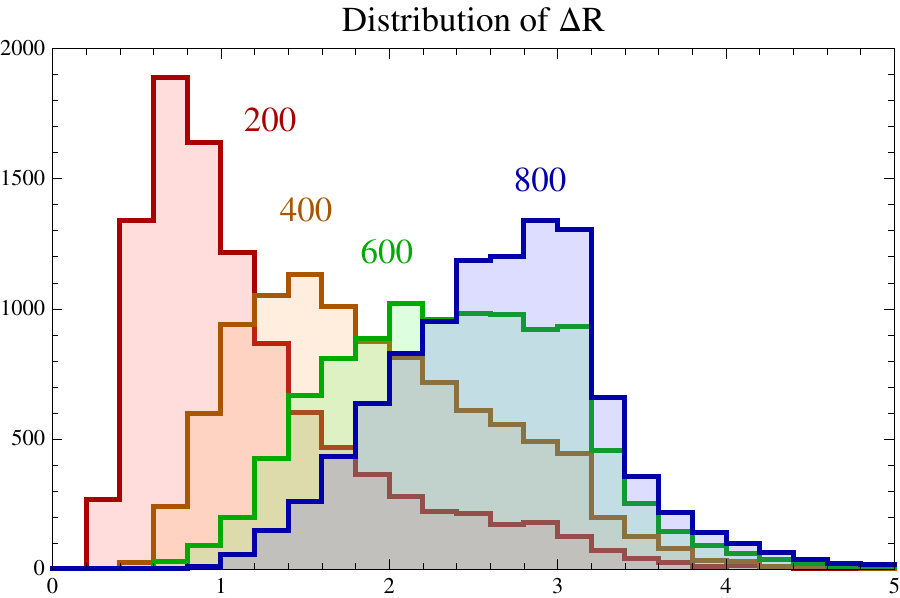}
}
\caption{Lab-frame angular separation between the two quarks from a stop decay. The stops are produced in the gluino cascade~\leqn{cascade}, following gluino pair-production at a 14 TeV LHC. We assume $\mg =1.2$ TeV, and vary the stop mass: $\mt=200, 400, 600$ and $800$ GeV distributions are shown in red, orange, green and blue, respectively. The distributions were calculated using {\tt MadGraph 5}~\cite{MG}.}
\label{fig:DeltaR}
\end{center}
\end{figure}

The key observation is that in a large section of the available parameter space, the stops produced in the gluino decays are relativistic. The stop boost in the gluino rest frame is given by
\beq
\gamma = \frac{1}{\sqrt{1-\beta^2}} = \frac{\mg^2 + \mt^2-m_t^2}{2\mg\mt}\,.
\eeq{stopvel}
so that stops are relativistic when $\mg\gg\mt$. For example, $\mg=1.2$ TeV and $\mt=200$ GeV yields $\beta\approx0.9$. 
Since gluinos themselves are mostly produced with non-relativistic speeds in the lab frame, such stops are typically also relativistic in the lab frame. In this regime, the two quarks produced in the stop decay are boosted in the same direction and have a small angular separation as can be seen in Fig.~\ref{fig:DeltaR}. The showers produced by the neighboring quarks tend to be merged into a single jet. Such ``stoppy" jets can be distinguished from regular QCD jets, as we will discuss in detail below, giving an extra handle that can be used to suppress the background and improve the search reach.  

\begin{figure}[h!]
\begin{center}
\centerline {
\includegraphics[width=2.92in]{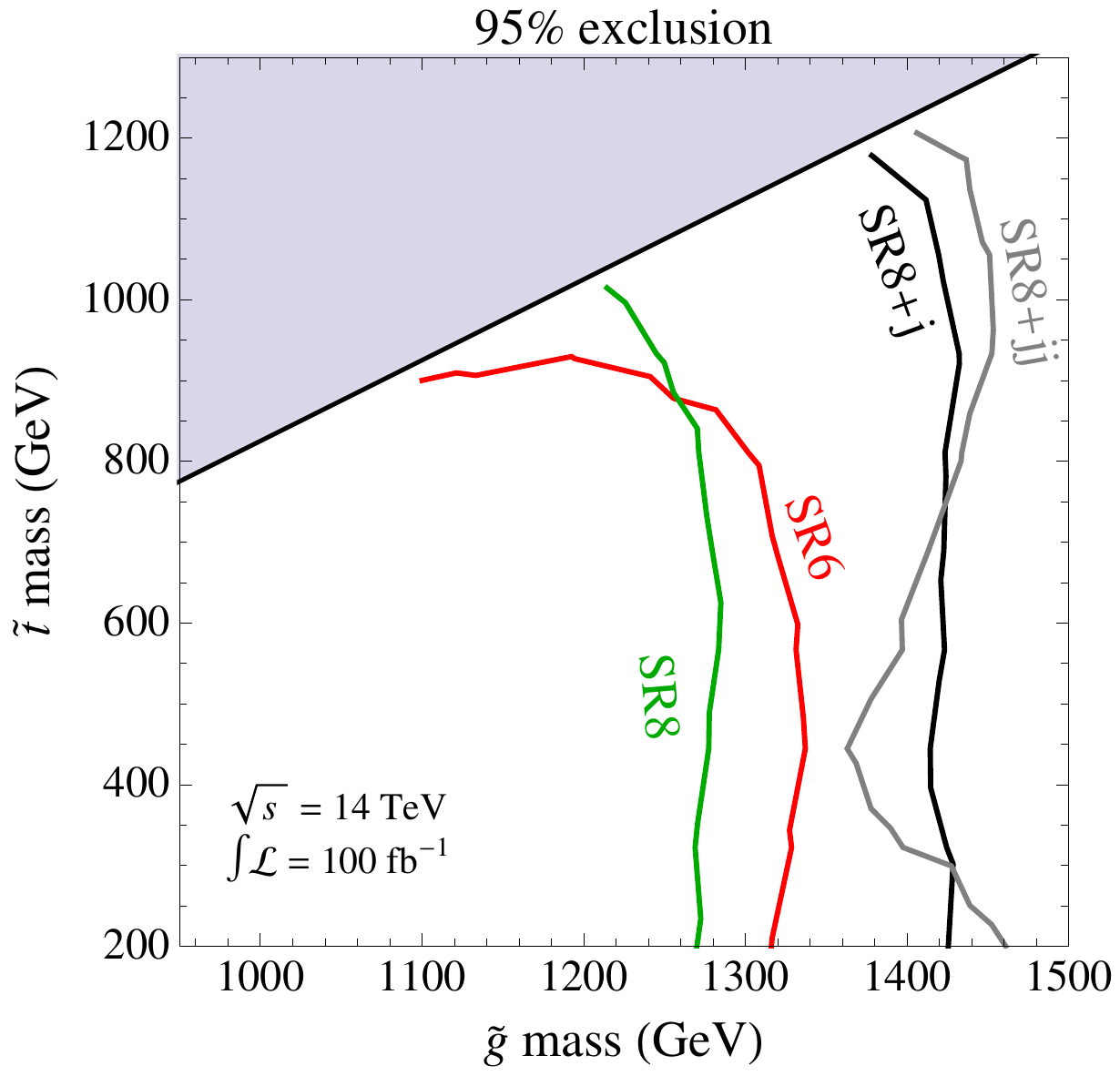}
\includegraphics[width=2.8in]{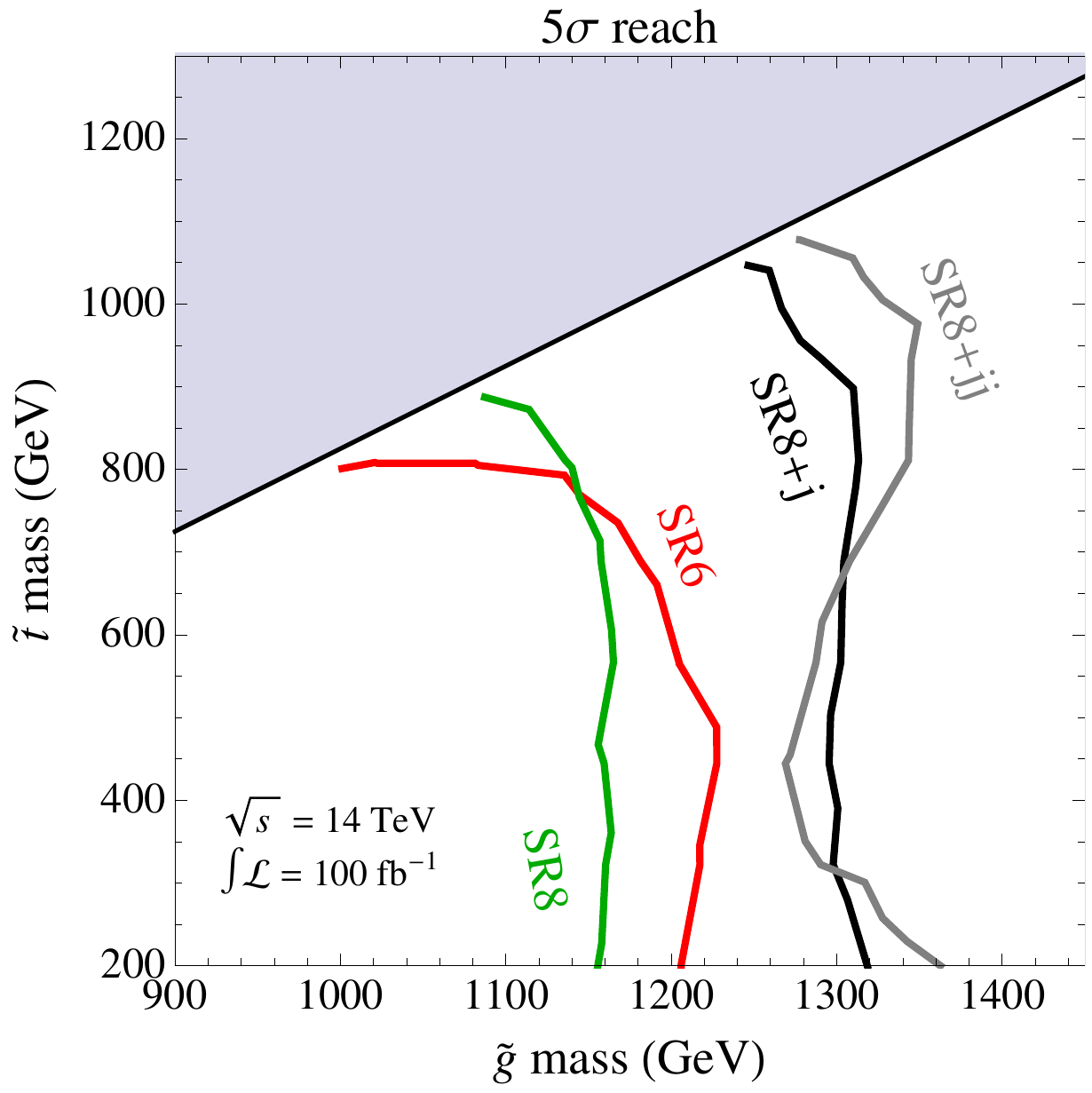}
}
\caption{Estimated 95\%~CL expected exclusion (left panel) and $5\sigma$ expected discovery (right panel) reach in the RPV SUSY simplified model parameter space at the 14 TeV LHC with 100 fb$^{-1}$. Red/green lines: reach of the analysis identical to the one in Ref.~\cite{CMS3}, for signal regions SR6/SR8. Black/gray: reach of the analysis with the SR8 cuts and an additional requirement of one/two jets with $M_j>175$ GeV. In the gray shaded region, the decay $\go\to\st t$ is kinematically forbidden.}
\label{fig:reach1}
\end{center}
\end{figure}

To assess the potential improvement, we performed a Monte Carlo study for the 14 TeV LHC. For this study, we simulated the signal, $pp\to \go\go$, using {\tt Pythia 8.162}~\cite{pythia}, for a large set of $(\mg, \mt)$ points. The leading order (LO) cross section provided by {\tt Pythia} is multiplied by the NLO K-factor for normalization. Gluino, top and $W$ decays are also treated in {\tt Pythia}, as are QCD initial radiation, showering and hadronization. Jet reconstruction is modeled with {\tt FastJet}~\cite{FastJet} using the anti-$k_T$ clustering algorithm. The dominant irreducible backgrounds, $t\bar{t}W$ and $t\bar{t}Z$, were simulated using the same tools. The cross sections for these processes are also normalized with NLO K-factors~\cite{BG_NLO}. 

	\begin{table}[h]
\begin{center}
		\tabcolsep 2.7pt
		\begin{scriptsize}
			\begin{tabular}{|r|c||c|c||c|c||c|c|}
			\hline
			process~~~~  &   $\sigma$(total)     & Eff(SR8)  & $\sigma$(SR8) &  Eff(1HMJ)  & $\sigma$  (SR8+1HMJ)                  &     Eff(2HMJ)                &       $\sigma$  (SR8+2HMJ)    \\
          \hline
          signal $(1200, 200)$ & 113 & 0.41 & 0.46 & 86 & 0.40 & 40 & 0.18 \\
          $(1200, 500)$ & 114 & 0.44 & 0.50 & 64 & 0.32 & 24 & 0.12 \\ 
          $(1200, 800)$ & 114 & 0.45 & 0.52 & 70 & 0.36 & 31 & 0.16 \\ 
          $(1300, 200)$ & 63 & 0.36 & 0.23 & 89 & 0.20 & 40 & 0.09 \\ 
          $(1300, 500)$ & 63 & 0.48 & 0.30 & 71 & 0.22 & 22 & 0.07 \\ 
          $(1300, 800)$ & 63 & 0.45 & 0.28 & 75 & 0.21 & 31 & 0.09 \\ 
          $(1300, 1100)$ & 62 & 0.30 & 0.19 & 81 & 0.15 & 43 & 0.08 \\ 
          $(1400, 200)$ & 35 & 0.39 & 0.14 & 95 & 0.13 & 48 & 0.07 \\ 
          $(1400, 500)$ & 35 & 0.44 & 0.15 & 73 & 0.11 & 27 & 0.04 \\ 
          $(1400, 800)$ & 35 & 0.43 & 0.15 & 78 & 0.12 & 34 & 0.05 \\ 
          $(1400, 1000)$ & 35 & 0.45 & 0.16 & 81 & 0.13 & 43 & 0.07 \\ 
          $(1400, 1200)$ & 35 & 0.29 & 0.1 & 80 & 0.08 & 40 & 0.04 \\ 
	\hline
background~~$t\bar{t}W$ & 590 & 0.07 & 0.38 & 4.7 & 0.02 & 0.3 & 0.001 \\
 $t\bar{t}Z$ & 910 & 0.03 & 0.30 & 7.9 & 0.02 & 0.6 & 0.002 \\
	\hline
			\end{tabular}
		\end{scriptsize}
\caption{Cross sections (in fb) and efficiencies (in \%) of signal and background processes, at the 14 TeV LHC. The signal points are labeled by $(\mg, \mt)$, both in GeV. The selection cuts are labeled as follows: SR8 refers to the cuts imposed by the CMS analysis~\cite{CMS3} in signal region 8 (see Table~\ref{tab:fromCMS}); 1HMJ means requiring at least one ``high-mass" jet ($M_j>175$ GeV); similarly, 2HMJ requires at least 2 jets with $M_j>175$ GeV. The 1HMJ and 2HMJ cuts are applied to the events that pass all SR8 cuts.}
\label{tab:cutflow}
\end{center}
	\end{table}

To set a benchmark point against which improvements can be judged, we estimated the reach of the searches currently performed by CMS~\cite{CMS3} at the 14 TeV LHC with $L_{\rm int}=100$ fb$^{-1}$. For this estimate, we implemented the cuts corresponding to the CMS signal regions listed in Table~\ref{tab:fromCMS} (with the exception of SR7, which would require a separate analysis due to an additional $b$-tagged jet requirement) on both signal and background samples. We modeled $b$-tagging by applying a $p_T$-dependent tagging efficiency for the CSVM tagger~\cite{CMSb} to all the jets that can be traced back to a $b$-hadron. The cut efficiencies for the signal and the background are listed in Table~\ref{tab:cutflow}. We then estimated the instrumental background. The two dominant sources are ``fake leptons" from sources such as heavy-flavor decays and misidentified hadrons, and ``charge flips", events with opposite-sign leptons where one of the charges is mismeasured. The ratio of the instrumental background to the irreducible component reported in Ref.~\cite{CMS3} is roughly between 1:1 and 2:1, depending on the signal region. This indicates that instrumental backgrounds will play an important role at 14 TeV as well. Unfortunately, detailed modeling of these backgrounds requires either detector simulation or data-based techniques. However, a rough estimate may be obtained as follows. Since the physical process primarily responsible for the instrumental backgrounds is top pair-production\footnote{We are grateful to Frank Wuerthwein for clarifying this point.}, it is reasonable to expect that the rates  scale approximately with the total $t\bar{t}$ cross section when the collision energy is increased from 8 to 14 TeV. Using this scaling and the instrumental background rates in various signal regions quoted in Ref.~\cite{CMS3}, we obtained corresponding estimates at 14 TeV. We found that the irreducible and instrumental background components scale by similar factors when going to 14 TeV: for example, our estimate of the instrumental/irreducible ratio at 14 TeV for the signal region SR6 is $0.86$, while for SR8 it is $1.62$, quite close to the ratios at 8 TeV. 

Combining the irreducible and instrumental backgrounds, we computed the exclusion levels expected under the assumption that the data exactly matches the background prediction, as well as the discovery reach defined by requiring at least a $5\sigma$ difference between  the signal+background and background-only predictions. The estimated exclusion and discovery reach contours are shown in Fig.~\ref{fig:reach1} for the two most sensitive signal regions: SR6 (red contour) and SR8 (green contour). 

\begin{figure}[tb]
\begin{center}
\centerline {
\includegraphics[width=3in]{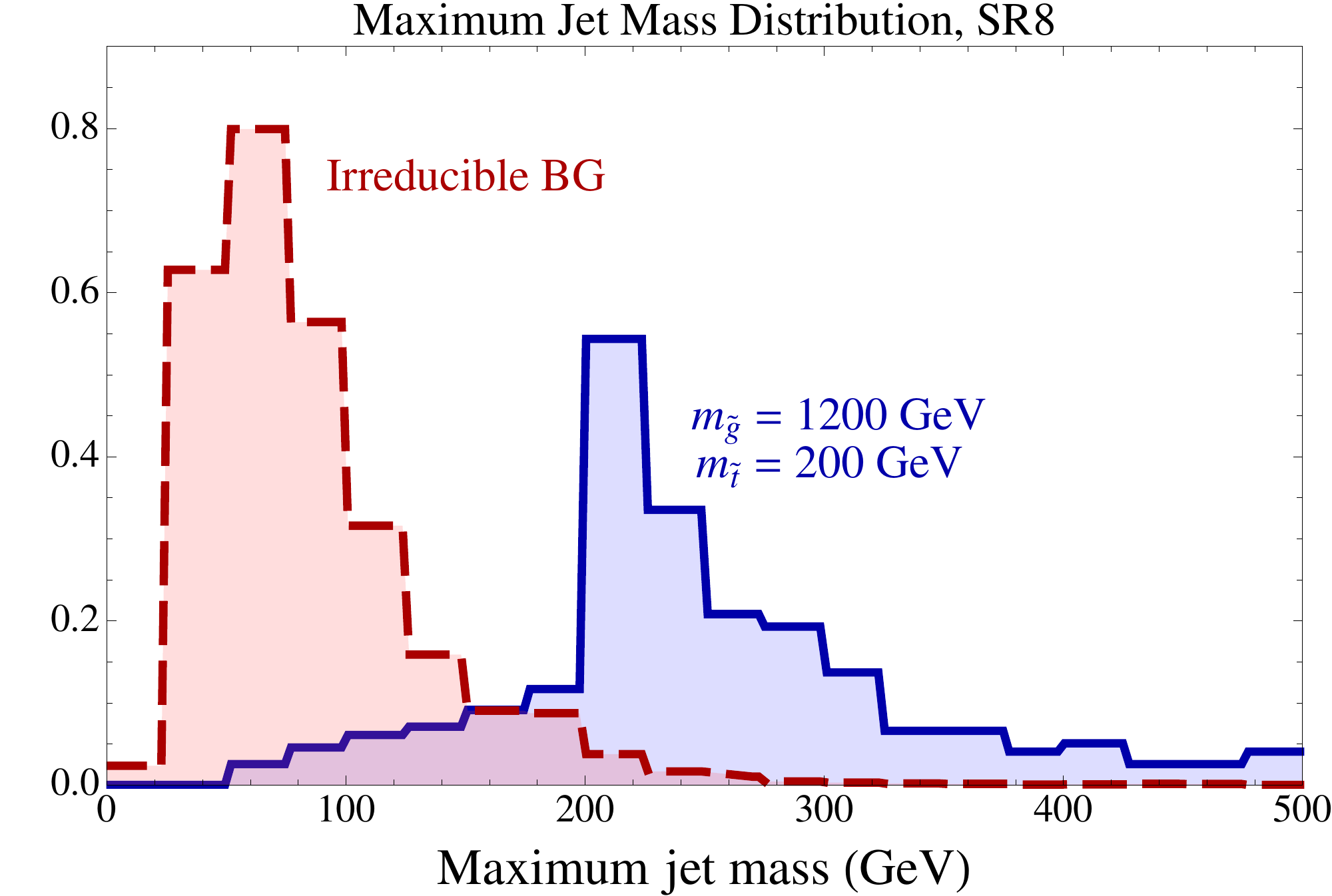}
\includegraphics[width=3in]{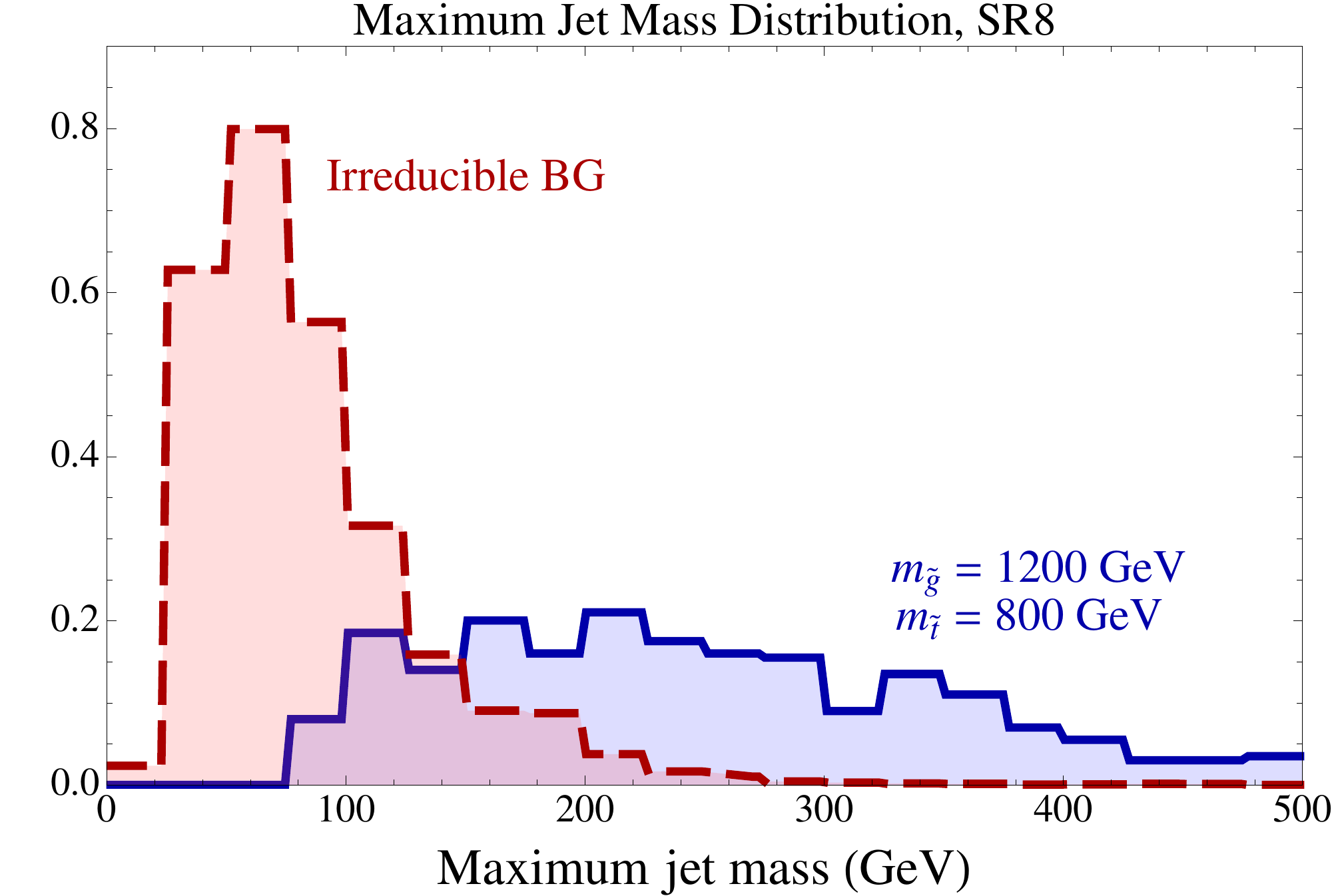}
}
\caption{Distributions of the largest jet invariant mass $M_j^{\rm max}$, in the signal (blue) and irreducible background (red) events passing SR8 cuts at the 14 TeV LHC. The signal is simulated for $(\mg, \mt)=(1200, 200)$ GeV (left panel) and $(1200, 800)$ GeV (right panel). The background includes the SM $t\bar{t}W$ and $t\bar{t}Z$ processes.}
\label{fig:Mjet}
\end{center}
\end{figure}

To identify the merged jets from stop decays, we first reclustered the samples, setting the jet opening angle to $\Delta R = 1.0$, as 
opposed to $\Delta R = 0.5$ used by the CMS analysis. Such ``fat" jets are already being used by experimental analyses involving jet substructure (see, for example, Refs.~\cite{CMSfat,ATLASfat}). We then computed the invariant mass $M_j$ of each jet. The distributions of the largest $M_j$ in each event, for both the signal and the (irreducible) background samples, are shown in Fig.~\ref{fig:Mjet}. It is obvious that $M_j^{\rm max}$ is an excellent signal/background discriminator. For the case $\mg\gg\mt$, illustrated in the left panel of the figure, the reason is obvious: the high-mass jets in the signal are due to boosted stop decays, and their masses peak around $\mt$. However, somewhat more surprisingly, this discriminator continues to work well in the regime $\mg\sim\mt$, as illustrated by the right panel of the figure. The reason for this is simply the large jet multiplicity in the signal, which at parton level has 6 quarks in the final state. In this situation, two independent parton showers (from different stops, or from a stop and a top) often get accidentally merged into a single jet which is more likely to have a large invariant mass than a single-parton QCD jet. (This phenomenon was previously noticed in~\cite{Hook}.) As a result, requiring massive jet(s) improves the reach of the search throughout the parameter space, and not just for large $\mg/\mt$.  

The improvement of the reach with the jet mass cut is shown by the black and gray lines in Fig.~\ref{fig:reach1}. This analysis imposes all of the SR8 cuts with the additional requirement of at least one or two high-mass jets with $M_j>175$ GeV. The efficiencies of these cuts, and cross sections after all cuts, are listed in Table~\ref{tab:cutflow}. For the reach estimate, we assumed that the efficiency of the jet invariant mass cuts on the instrumental and irreducible backgrounds are the same (which seems reasonable since both contain QCD jets of similar energies). We found that gluinos up to $1.4-1.45$ TeV can be excluded at the 95\%~CL, while gluinos up to $1.3-1.35$ TeV can be discovered at the 5$\sigma$ level at the 14 TeV LHC with $100$ fb$^{-1}$. The dependence of the reach on the stop mass is quite weak, especially when the analyses with $\geq 1$ and $\geq 2$ high-mass jets are combined.  
 
\begin{figure}[tb]
\begin{center}
\centerline {
\includegraphics[width=3in]{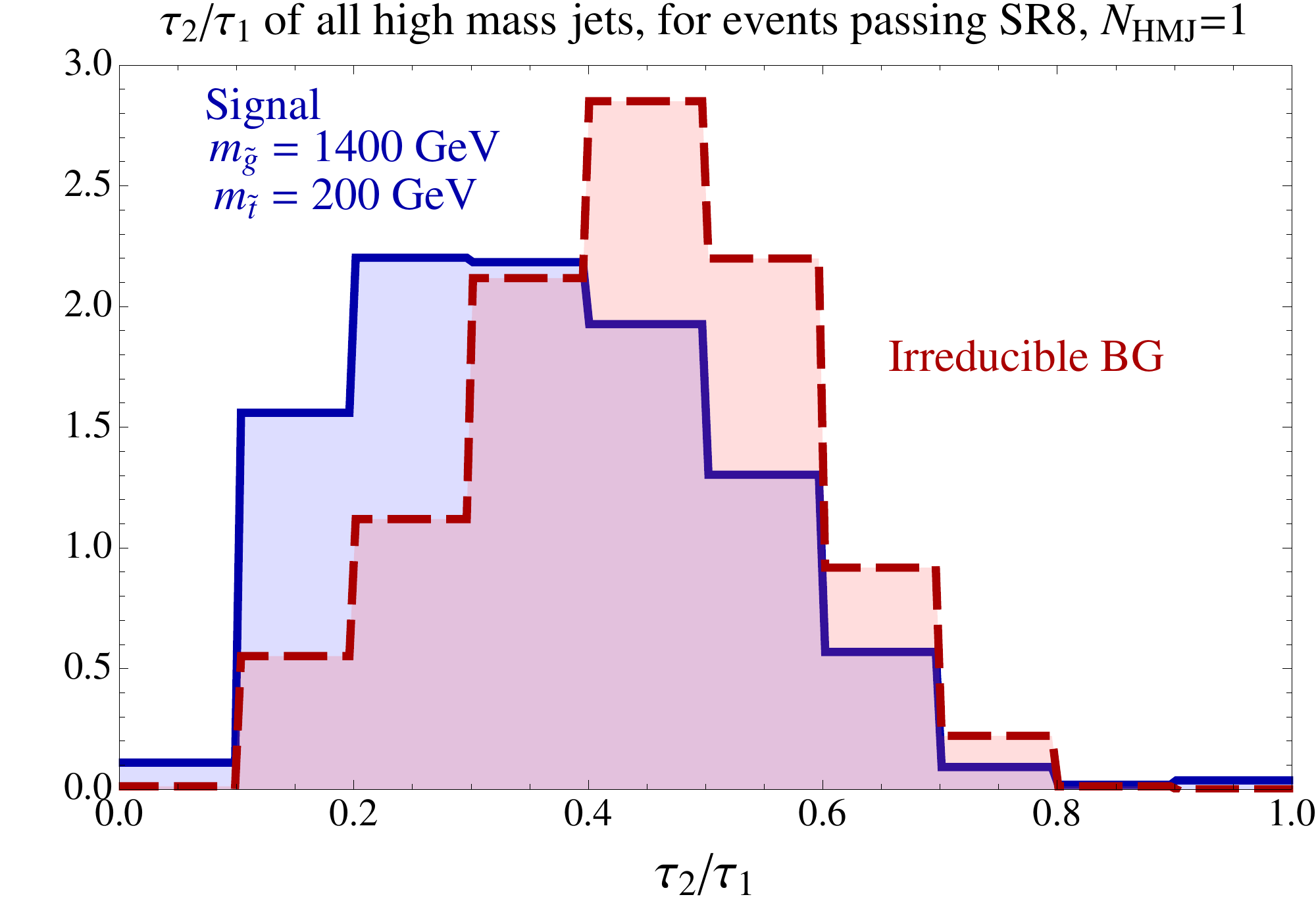}
\includegraphics[width=3in]{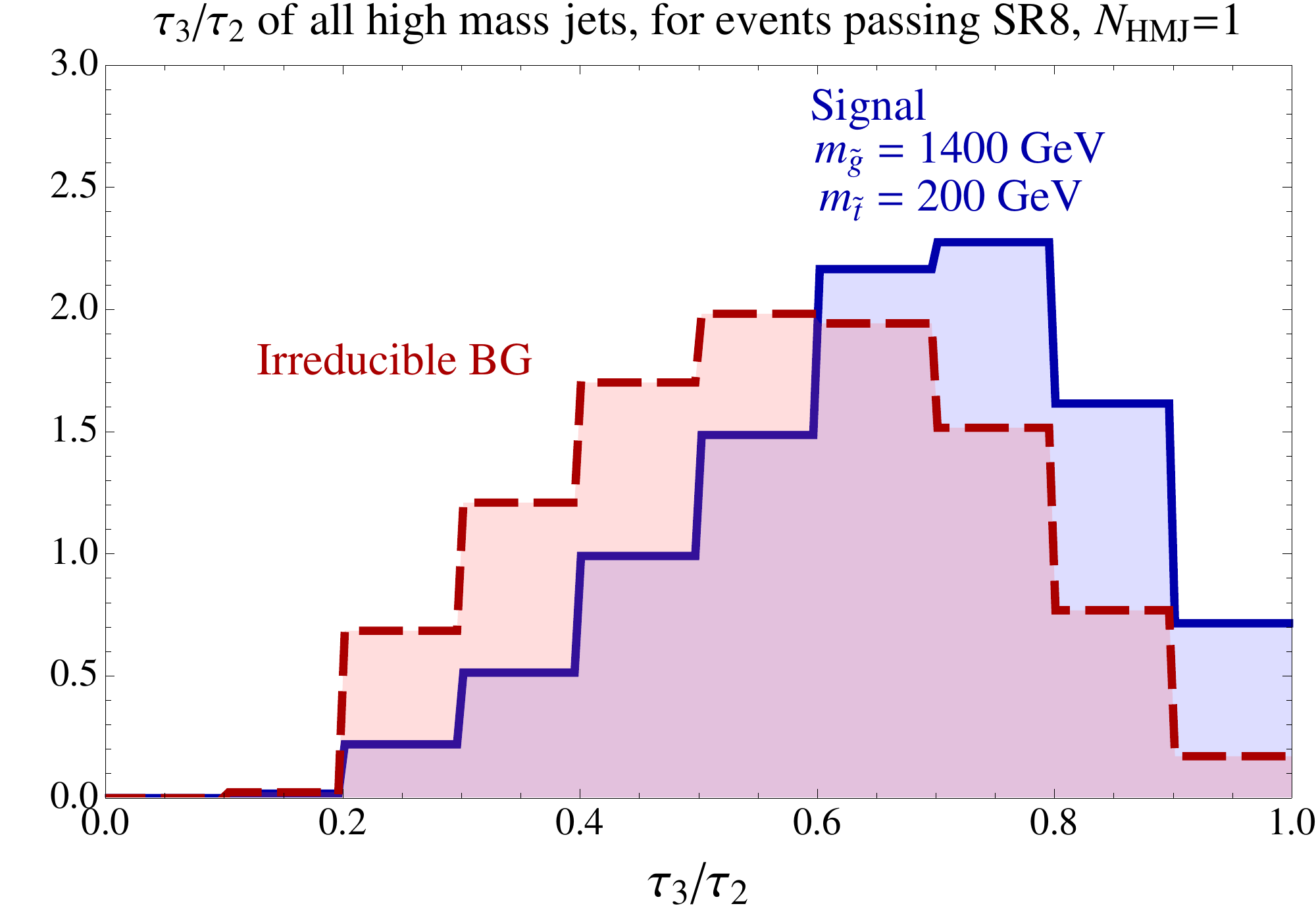}
}
\caption{Distributions of $N$-subjettiness observables, $\tau_2/\tau_1$ (left) and $\tau_3/\tau_2$ (right), for the high-mass jets ($M_j>175$ GeV) in the signal (blue) and irreducible background (red) events passing SR8 cuts. The signal is simulated for $(\mg, \mt)=(1400, 200)$ GeV. All distributions are normalized to unit area.}
\label{fig:Nsubjet}
\end{center}
\end{figure}

An even stronger separation  of signal and background can be achieved by noticing that the high-mass jets in the background are primarily due to boosted, fully hadronic tops. Such jets have three hard partons. In contrast, the signal jets typically have two hard partons from a two-body stop decay. To exploit this, we used the $N$-subjettiness technique proposed by Thaler and 
Van Tilburg~\cite{Nsub}. In this approach, observables $\tau_N$ are defined with $N=1, 2, \ldots$. A low value of the ratio $\tau_N/\tau_{N-1}$ indicates that the jet likely has an $N$-pronged substructure. For example, the distributions of jets with $M_j>175$ GeV in $\tau_2/\tau_1$ and $\tau_3/\tau_2$ observables are shown in Fig.~\ref{fig:Nsubjet}, where in the signal simulation we assumed $(\mg, \mt)=(1400, 200)$ GeV, and 
used the {\tt onepass\_kt\_axes} minimization scheme and $\beta=1.1$. As expected, low values of $\tau_2/\tau_1$ are favored in the signal, while low values of $\tau_3/\tau_2$ are favored in the background. It should be noted that with the 100 fb$^{-1}$ data set, the reach of the jet-mass based searches shown in Fig.~\ref{fig:reach1} is already statistics-limited, so no further improvement can be achieved by cutting on the $N$-subjettiness observables. However, they can be useful for larger data sets, or as a part of more globally optimized set of cuts.

\begin{figure}[tb]
\begin{center}
\centerline {
\includegraphics[width=2.9in]{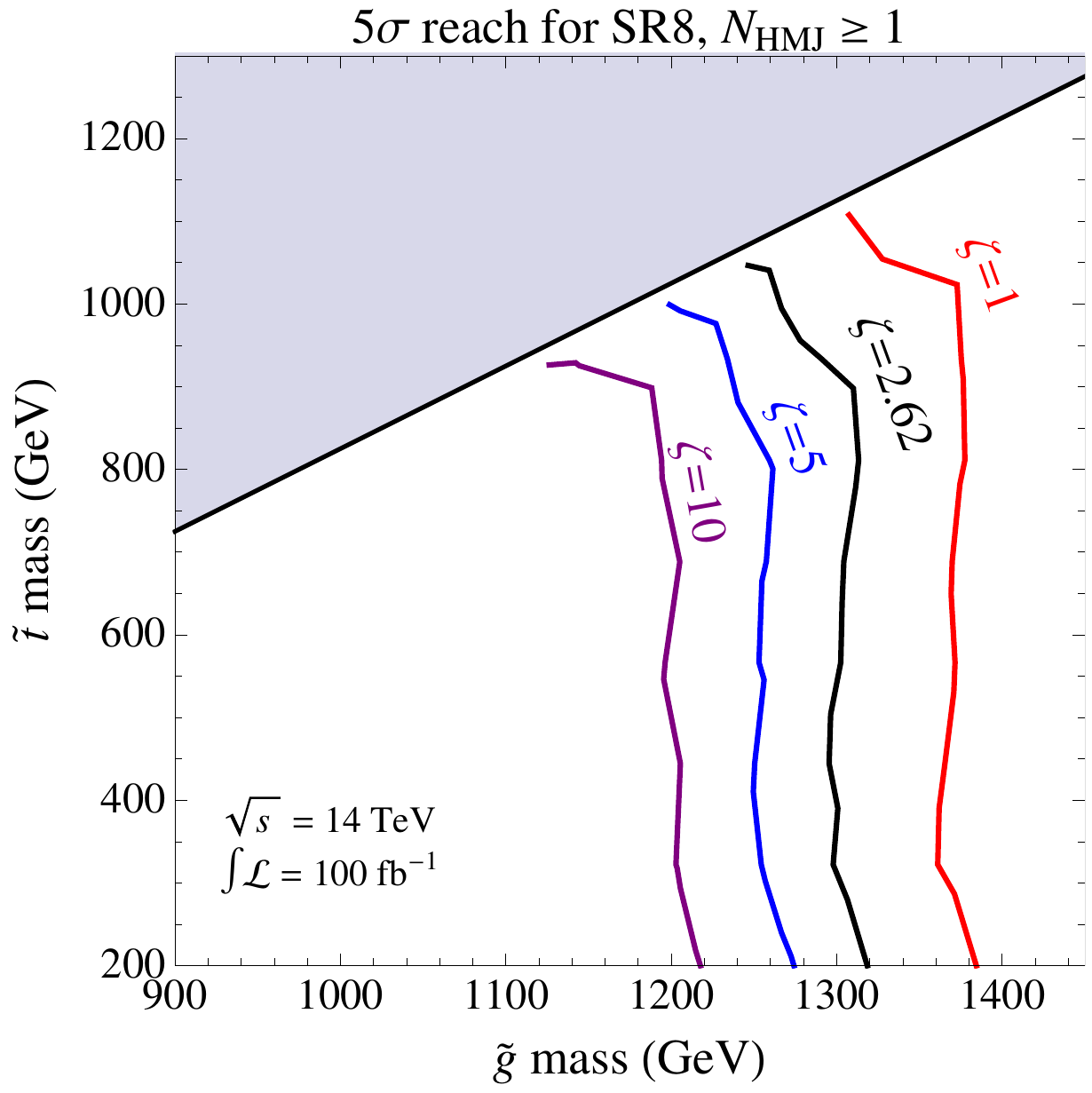}
\includegraphics[width=2.9in]{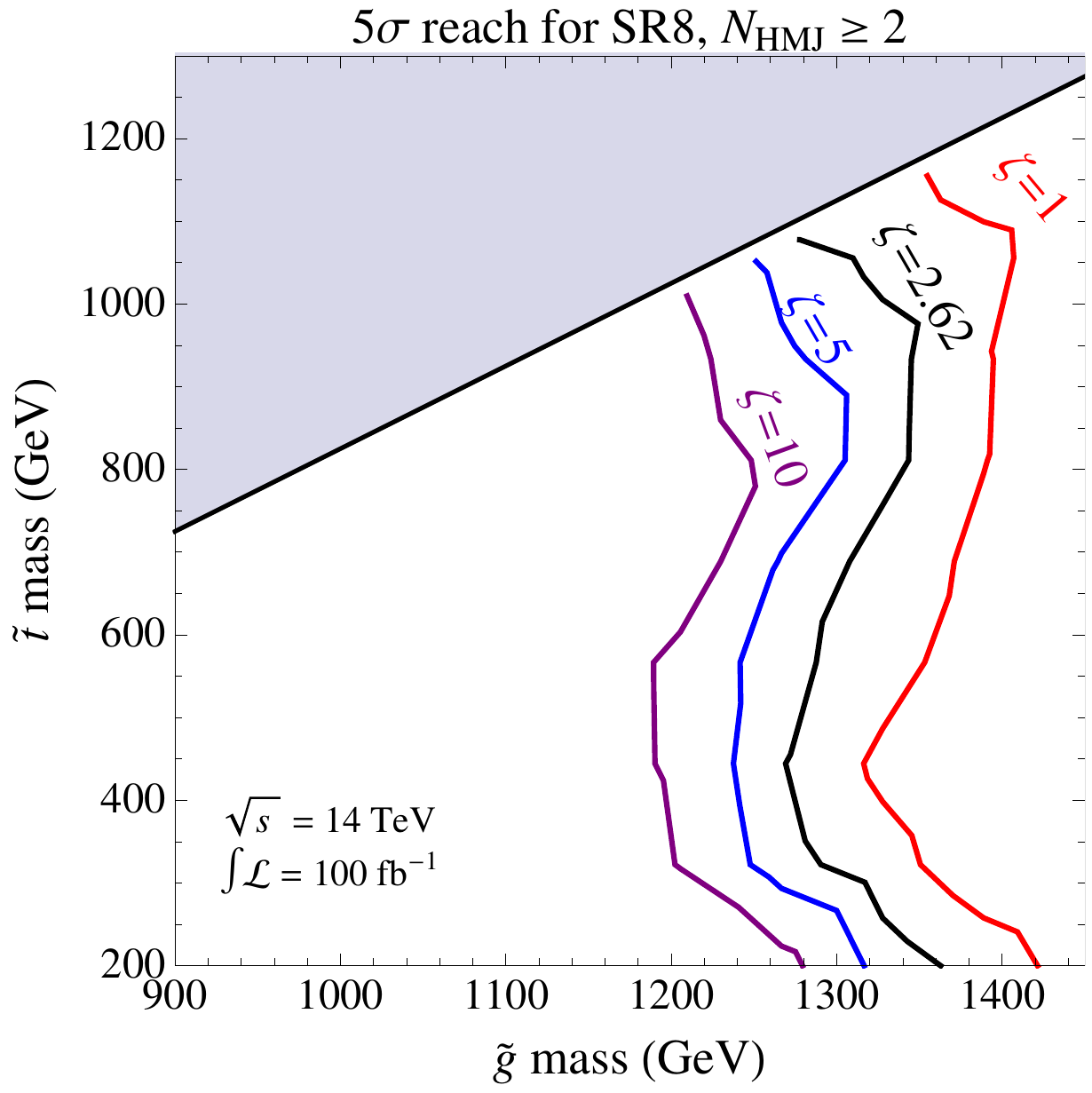}
}
\caption{Estimated discovery reach in the RPV SUSY simplified model parameter space, at the 14 TeV LHC with 100 fb$^{-1}$ of data, for a range of assumptions concerning the instrumental background. The selection cuts are SR8, plus $\geq1$ (left) or $\geq2$ (right) jets with $M_j>175$ GeV. The value $\zeta=2.62$ is the estimate obtained by rescaling from 8 TeV and used in Fig.~\ref{fig:reach1}. In the gray shaded region, the decay $\go\to\st t$ is kinematically forbidden.}
\label{fig:reach2}
\end{center}
\end{figure}

Since no detector simulation could be performed for this study, our instrumental background estimate is clearly very crude and has a large uncertainty. To illustrate how this uncertainty affects the reach of the proposed search, we define
\beq
\zeta = \frac{{\rm Total~BG~Rate}}{{\rm Irreducible~BG~Rate}}\,,
\eeq{zeta} 
where both rates include all the cuts imposed in a particular analysis. Fig.~\ref{fig:reach2} shows the variation of the reach for values of $\zeta$ between 1 and 10, for the same analysis as in Fig.~\ref{fig:reach1} (SR8 plus $\geq1$ or $\geq2$ jets with $M_j>175$ GeV). The reach estimates are relatively robust with respect to the uncertainty in the instrumental background estimate, due to a strong dependence of the signal rates on $\mg$.    

\section{Discussion and Conclusions}
\label{sec:conc}

The main results of this paper can be summarized as follows:

\begin{itemize}

\item The current CMS searches for anomalous events with SSDL and $b$-jets place a lower bound of about 800 GeV on the gluino mass in the gluino-stop simplified model of RPV/MFV SUSY. The bound is only weakly sensitive to the stop mass, as long as an on-shell decay $\go\to\st t$ is kinematically allowed. 

\item A search identical to the current CMS search, implemented at the 14 TeV LHC with 100 fb$^{-1}$ of data, is estimated to have the sensitivity to exclude gluino masses up to about 1.3 TeV at the 95\% CL, and a $5\sigma$ discovery reach of about 1.2 TeV. Again, these are largely insensitive to the stop mass.

\item An addition of a cut on the jet invariant mass improves the 95\% CL exclusion reach and the $5\sigma$ discovery reach to approximately 1.45 TeV and 1.35 TeV, respectively. While the improvement in terms of the gluino mass is only about 10\% in both cases, it is still very significant since the gluino cross section drops very rapidly with mass. 

\end{itemize}

While the motivation for our analysis comes primarily from the MFV SUSY model~\cite{CGH}, the results apply quite generally to RPV models with a stop LSP, decaying via a $UDD$-type operator. (See, for example, Ref.~\cite{UDD} for a recent discussion of such models.) A non-MFV flavor structure of the stop decay operator may result in fewer $b$-jets, but since top quarks still provide two genuine $b$-jets per event, even in this case the efficiencies of the cuts should not be strongly degraded. 

For our signature to work, it is crucial that the gluino be a  Majorana particle. If the gluino is Dirac, no SSDL signature is possible, and other handles must be used to suppress the SM background. However, high-mass jets from stop decays are still present in this situation, and can provide a useful discriminant~\cite{Harvard}. It would be interesting to see if, in addition to stop jets, massive jets formed by the boosted SM tops produced from the same gluino decays can be useful in this context. (The utility of boosted top-jets in searching for the gluino-stop cascade decays in R-parity conserving SUSY has been pointed out in~\cite{us}.) We leave this possibility for future study.

\vskip0.8cm
\noindent{\large \bf Acknowledgments} 
\vskip0.3cm

We would like to thank Didar Dobur, Gala Nicolas-Kaufman, Werner Sun, Jordan Tucker, Luke Winstrom, Frank Wuerthwein and Felix Yu. This research is supported by the U.S. National Science Foundation through grant PHY-0757868 and CAREER grant PHY-0844667.

\begin{appendix}

\section{Details of the Recasting Procedure}
\label{app:recast}

To recast the CMS SSDL+MET+$b$ analysis in terms of the RPV SUSY model, we follow closely the instructions provided by  CMS in~\cite{CMS3} and its predecessors~\cite{SUS-11-020, SUS-12-017}. The only significant difference is in the treatment of leptons. The instructions recommend analyzing leptons at parton level, by taking the leptons that pass the kinematic cuts and applying the selection efficiencies given in Section 7 of~\cite{CMS3}.  These selection efficiencies, which account for lepton identification efficiencies, isolation cuts, and detector effects, had been computed from Monte Carlo studies of simplified model A1 ($p p \rightarrow \tilde{g} \tilde{g}$, $\tilde{g} \rightarrow t\overline{t}\tilde{\chi}^0$) at the RPC SUSY benchmark point LM9.  However, because the leptons in the RPV SUSY signal process may come from boosted tops, there is extra hadronic activity near the leptons, and the LM9 selection efficiencies do not properly model the isolation cuts for the RPV signal. Therefore, we extract the isolation cut efficiencies for RPV from our signal MC. To do so, we impose a lepton isolation cut on the {\it hadronized} signal MC events. Following~\cite{SUS-11-020}, $Iso(\hat{\ell})$ is defined as  a scalar sum of the lepton $p_T$'s and photon and hadron $E_T$'s within a cone of size $\Delta R \equiv \sqrt{(\Delta\eta)^2 + (\Delta\phi)^2} < 0.3$ about the lepton, not including the $p_T$ of the lepton itself:
\beq
Iso(\hat{\ell}) \equiv \frac{\sum_{\Delta R < 0.3} p_T (\ell \neq \hat{\ell}) + \sum_{\Delta R < 0.3} E_T (\gamma) + \sum_{\Delta R < 0.3} E_T (h)}{p_T (\hat{\ell})}.
\eeq{Iso}
To pass the isolation cut, the lepton must have have $Iso(\hat{\ell}) < 0.1$. On top of the isolation cut, we impose the identification efficiency, which we assume to be independent of $p_T$, $\eta$, and the physical process: 73\% for electrons and 84\% for muons. The identification efficiency for each lepton species is extracted by simulating the A1 LM9 benchmark model at hadron level, computing the lepton isolation cut efficiency Eff$(Iso)$ for this sample using Eq.~\leqn{Iso}, and dividing the total selection efficiency reported by CMS by Eff$(Iso)$.

The rest of the lepton analysis emulates~\cite{CMS3} as closely as possible.  From the set of selected leptons, we choose the ``SSDL pair'': the same-sign pair with the highest $p_T$ and a pair invariant mass of at least 8 GeV.  We then apply the dilepton trigger efficiency: 96\% for $ee$, 93\% for $e\mu$, and 88\% for $\mu\mu$.  We veto events where a third lepton (with $p_T > 10$ GeV, the normal $|\eta|$ cuts, and $Iso(\l_3) < 0.2$) forms an opposite-sign same-flavor pair with one of the SSDL pair leptons, with a pair invariant mass between 76 and 106 GeV.  We also veto events where a third lepton (with $p_T > 5$ GeV, the normal $|\eta|$ cuts, and $Iso(\l_3) < 0.2$) forms an opposite-sign same-flavor pair with one of the SSDL pair leptons, with a pair invariant mass below 12 GeV.

The remaining physics objects are handled at parton level, following the instructions.  The number of jets is a count of colored partons passing the kinematic cuts: $p_T > 40$ GeV and $|\eta| < 2.4$.  To count $b$-tagged jets, we apply a $p_T$-dependent tagging efficiency, parameterized in Section 7 of~\cite{CMS3}, to all the $b$ quarks that pass the jet kinematic cuts.  To implement the cuts on $H_T$ and $\slashed{E}_T$, we compute ``generator-level'' quantities gen-$H_T$ and gen-$\slashed{E}_T$, and use the turn-on efficiency curves parameterized in Section 7 of~\cite{SUS-11-020} to get efficiencies for the cuts.  gen-$H_T$ is the scalar sum of $p_T$'s of the jets that pass the kinematic cuts, and gen-$\slashed{E}_T$ is the magnitude of the vector sum of the $\vec{p}_T$'s of non-interacting final-state particles.

\end{appendix}

\end{document}